\documentclass[10pt,twocolumn,letterpaper]{article}
\usepackage[table,xcdraw]{xcolor}
\usepackage{caption}
\usepackage[utf8]{inputenc}
\usepackage{algorithm}
\usepackage{algpseudocode}
\usepackage{amsmath} 
\usepackage{booktabs}
\usepackage{multirow}
\usepackage{subcaption}
\usepackage{graphicx}
\usepackage[a4paper, margin=1in]{geometry}
\usepackage{cvpr}              % To produce the CAMERA-READY version
\definecolor{cvprblue}{rgb}{0.21,0.49,0.74}
\usepackage[pagebackref,breaklinks,colorlinks,citecolor=cvprblue]{hyperref}

\def\paperID{*****} % *** Enter the Paper ID here
\def\confName{ICCV}
\def\confYear{2025}

\title{Multi Source COVID-19 Detection via Kernel-Density-based Slice Sampling}
\author{$^\ddagger$$^1$Chia-Ming Lee, $^1$Bo-Cheng Qiu, $^2$Ting-Yao Chen, $^1$Ming-Han Sun, $^2$Fang-Ying Lin \\$^1$Jung-Tse Tsai, $^2$I-An Tsai, $^1$Yu-Fan Lin,$^\dagger$$^{1,2}$Chih-Chung Hsu
\\
$^1$National Cheng Kung University \quad
$^2$National Yang Ming Chiao Tung University%\\
}

\begin{document}
\maketitle
 
\begin{abstract}
We present our solution for the Multi-Source COVID-19 Detection Challenge, which classifies chest CT scans from four distinct medical centers. To address multi-source variability, we employ the Spatial-Slice Feature Learning (SSFL) framework with Kernel-Density-based Slice Sampling (KDS). Our preprocessing pipeline combines lung region extraction, quality control, and adaptive slice sampling to select eight representative slices per scan. We compare EfficientNet and Swin Transformer architectures on the validation set. The EfficientNet model achieves an F1-score of 94.68, compared to the Swin Transformer's 93.34. The results show the performance of our KDS-based pipeline on multi-source data and highlight the importance of dataset balance in multi-institutional medical imaging evaluation.
\end{abstract}
\section{Introduction}

Deep learning has become a cornerstone of modern medical--image analysis, demonstrating strong predictive power across diverse healthcare applications~\cite{kollias2018deep,kollias2020deep}.  
During the COVID-19 pandemic, this momentum spurred rapid development of computer-aided diagnostic systems.  
Chest computed-tomography (CT) imaging, in particular, can reveal hallmark pulmonary manifestations such as ground-glass opacities and bilateral infiltrates that correlate with infection severity~\cite{kollias2021mia,kollias2022ai}.  
Early 3-D CNN pipelines trained on large, curated CT datasets confirmed that automated screening could approach radiologist-level sensitivity for COVID-19 detection and severity grading~\cite{arsenos2022large}.
 
Despite these successes, multi-institutional deployment remains difficult. CT volumes collected at different hospitals vary in scanner models, reconstruction kernels, slice spacing and patient demographics, introducing a domain shift that can markedly degrade performance when a network is evaluated outside its source domain~\cite{kollias2023deep}.  
Recent studies emphasise that robust COVID-19 screening must explicitly address site heterogeneity through domain adaptation, fairness and explainability mechanisms~\cite{kollias2023ai,kollias2024domain}.  
Transparent adaptation strategies---which align latent representations while exposing feature changes to clinicians---have been shown to mitigate such shifts without sacrificing interpretability~\cite{kollias2020transparent}.

To probe these issues, we participate in the \emph{PHAROS-AFE-AIMI Multi-Source Covid-19 Detection}, which provides chest CT scans from four distinct medical centres and requires binary classification of \textsc{covid} versus \textsc{non-covid} cases. The benchmark’s heterogeneous acquisition protocols mirror real-world deployment scenarios and thus constitute a demanding test of domain generalisation~\cite{arsenos2023data,gerogiannis2024covid}.
 
Our approach extends the SSFL framework of Hsu \emph{et al.}\cite{hsu2023} by integrating a \emph{Kernel-Density–based Slice-Sampling} (KDS) \cite{hsu2024} scheme that selects diagnostically relevant slices while preserving anatomical coverage.  
Because COVID-19 lesions are confined to lung parenchyma regardless of scanner vendor, we first extract precise lung masks with the recent vision–language segmentation pipeline \textsc{Sam2Clip2Sam}~\cite{kollias2024sam2clip2sam}.  
KDS then draws balanced slice subsets conditioned on kernel-density estimates of regional intensity distributions, yielding a compact yet domain-invariant representation for downstream classification.  
We compare a lightweight CNN (EfficientNet) and a transformer backbone (Swin Transformer) under identical preprocessing, demonstrating that careful slice selection narrows the gap between model sizes while improving cross-site robustness.

\section{Methodology}

\begin{figure}
    \centering
    \includegraphics[width=1\linewidth]{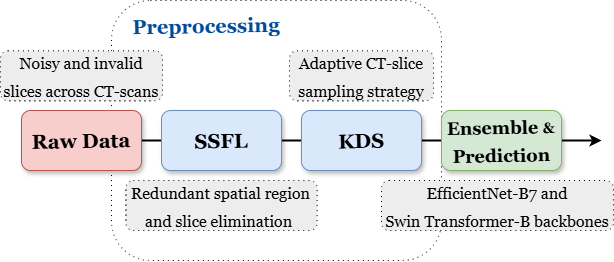}
    \caption{Pipeline overview of the proposed multi-source COVID-19 detection approach, showing the preprocessing steps (SSFL and KDS) and model architectures.
}
    \label{fig:overview}
\end{figure}

We present our solution for the Multi-Source COVID-19 Detection Challenge, which addresses the classification of chest CT scans across data from four distinct medical centers. Our approach builds upon the SSFL framework~\cite{hsu2024,hsu2023,hsu2023report,hsu2024report}, incorporating preprocessing and sampling techniques to handle multi-source variability. Figure~\ref{fig:overview} illustrates the overall architecture of our approach.

\subsection{Spatial-slice Feature Learning}

To address the heterogeneity across different medical centers, we implement a preprocessing pipeline for data consistency.

\textbf{Lung Region Extraction:} We first apply minimum filtering with a $3 \times 3$ kernel to the original CT images to smooth noise and suppress artifacts. Following this, fixed-threshold binarization with threshold $t$ and hole filling are performed to separate lung regions from the background. This step identifies diagnostically relevant content across different scanning protocols.

The filtered image $Z_{\text{filtered}}$ is obtained through:
\begin{equation}
Z_{\text{filtered}}(i, j) = \frac{\sum_{p=-k}^{k} \sum_{q=-k}^{k} w(p, q) \times Z(i + p, j + q)}{\sum_{p=-k}^{k} \sum_{q=-k}^{k} w(p, q)}
\end{equation}
where $w(p, q)$ represents the weight at position $(p, q)$ in the filter kernel with $k = 1$.
Afterwards, we perform consistency checks on image dimensions within each CT scan folder. Scans with inconsistent image sizes or containing fewer than 5 slices are excluded to maintain data integrity. This quality control step ensures that only complete scans are used for subsequent analysis.

\begin{figure}
    \centering
    \includegraphics[width=1\linewidth]{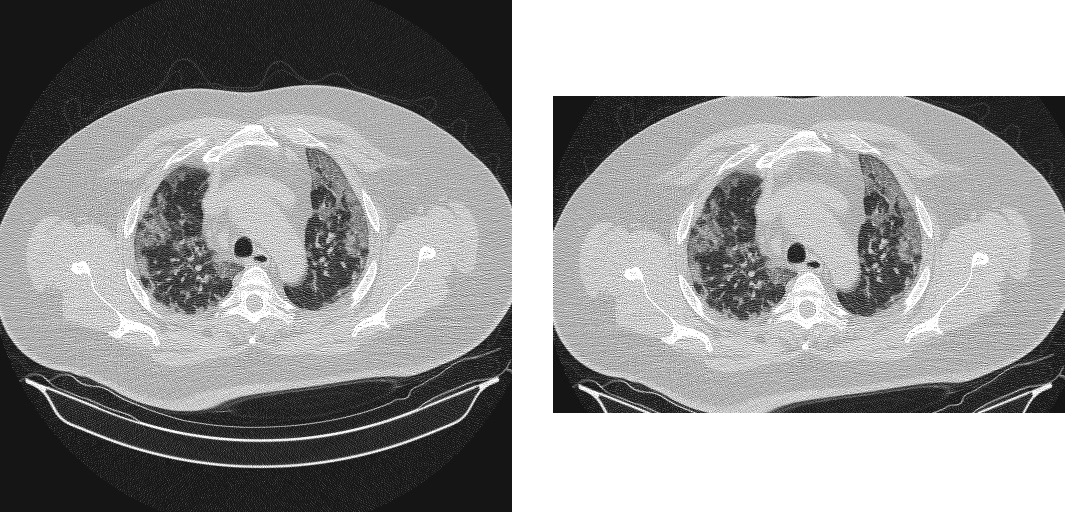}
    \caption{Example of a CT slice before and after lung region cropping. Left: original slice. Right: cropped slice.}
    \label{cropping.jpg}
\end{figure}

\textbf{Lung Region Cropping:} Based on the identified lung regions, we determine cropping boundaries using threshold-based segmentation with empirically assigned threshold $t$:
\begin{equation}
\text{Mask}[i, j] = \begin{cases}
0, & \text{if } Z_{\text{filter}}[i, j] < t \\
1, & \text{if } Z_{\text{filter}}[i, j] \geq t
\end{cases}
\end{equation}
This process removes background areas and focuses on lung tissue regions across different medical centers. All cropped images are resized to $256\times 256$ pixels for model input. Figure~\ref{cropping.jpg} shows an example of lung region cropping, demonstrating the focus on relevant areas after preprocessing.

\subsection{Kernel-Density-based Slice Sampling}

Following the identification of CT slice ranges, we apply our KDS method to select representative slices. This approach addresses varying scan lengths and slice quality across different sources.

\textbf{Distribution Estimation:} We apply Kernel Density Estimation (KDE) to estimate the distribution of lung areas within the slice range:
\begin{equation}
\hat{f}_h(x) = \frac{1}{sh} \sum_{i=1}^{s} K\left(\frac{x - x_i}{h}\right)
\end{equation}
where $h$ is the bandwidth constant calculated by Scott's rule with $h = 1.06 \cdot \sigma \cdot n^{-1/5}$, $s = 100$ for smoothing factor, and $K$ is a Gaussian kernel:
\begin{equation}
K(x, x') = \exp\left(-\frac{\|x - x'\|^2}{2\sigma^2}\right)
\end{equation}
where $\sigma^2 = h^2/2$ for the kernel bandwidth.

\begin{figure}
    \centering
    \includegraphics[width=1\linewidth]{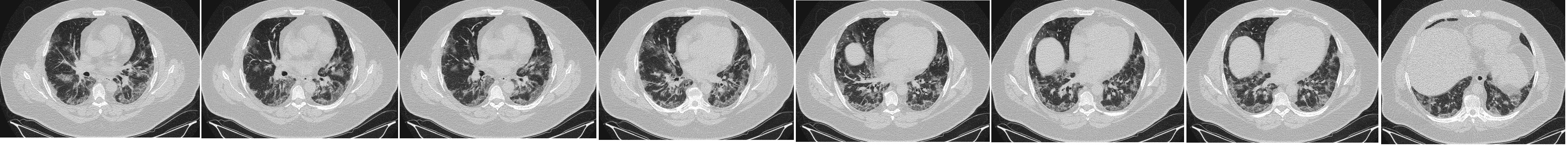}
    \caption{Example of representative CT slices selected using KDS.}
    \label{Sampling.png}
\end{figure}

\textbf{Adaptive Sampling:} The Cumulative Distribution Function (CDF) is calculated and partitioned into 8 percentile intervals:
\begin{equation}
F(x) = \int_{-\infty}^{x} \hat{f}_h(t)dt, \quad F(q_p) = p
\end{equation}
We sample exactly 8 representative slices per scan, with the number of slices from each interval proportional to its contribution to the overall CDF. This provides coverage of the lung anatomy while maintaining computational efficiency. Figure~\ref{Sampling.png} shows examples of CT slices selected through our KDS method.

\section{Experiment Results}
\label{sec:exp}

\subsection{Dataset Description}

The dataset used in this study was obtained from the multi-source COVID-19 Detection Challenge. These cases originate from four distinct source data centers, labeled numerically from 0 to 3. All CT scans were categorized into COVID-19 and non-COVID-19 classes.

The training set consists of 564 COVID-19 scans and 659 non-COVID-19 scans. The validation set included 128 COVID-19 scans and 180 non-COVID-19 scans. The test set contained 1,488 cases, for which the class labels were not provided and needed to be predicted. The overall dataset statistics are summarized in Table~\ref{tab:dataset_subgroup1_resizebox}.

\begin{table}[!ht]
  \centering
  \caption{Distribution of CT scan and slice counts across different data sources and COVID-19 status for training, validation, and testing sets.}
  \label{tab:dataset_subgroup1_resizebox}
  
  \resizebox{0.5\textwidth}{!}{%
    \begin{minipage}{0.55\textwidth}
    
      % --- 子表格 1: Scan-level ---
      \begin{subtable}[t]{\textwidth}
        \centering
        \caption{Scan-level counts}
        \label{subtab:scan_counts_resizebox}
        \begin{tabular}{lcrrr}
          \toprule
          \textbf{Type} & \textbf{Source} & \textbf{COVID} & \textbf{non-COVID} & \textbf{Total} \\
          \midrule
          \multirow{4}{*}{Training}
            & 0 & 175 & 164 & 339 \\
            & 1 & 175 & 165 & 340 \\
            & 2 &  39 & 165 & 204 \\
            & 3 & 175 & 165 & 340 \\
          \midrule
          \multirow{4}{*}{Validation}
            & 0 &  43 &  45 &  88 \\
            & 1 &  43 &  45 &  88 \\
            & 2 &   0 &  45 &  45 \\
            & 3 &  42 &  45 &  87 \\
          \midrule
          \multirow{4}{*}{Total}
            & 0 & 218 & 209 & 427 \\
            & 1 & 218 & 210 & 428 \\
            & 2 &  39 & 210 & 249 \\
            & 3 & 217 & 210 & 427 \\
          \midrule
          Testing & -- & -- & -- & 1,488 \\
          \bottomrule
        \end{tabular}
      \end{subtable}
      
      \vspace{1.5em} % 垂直間距
      
      % --- 子表格 2: Slice-level ---
      \begin{subtable}[t]{\textwidth}
        \centering
        \caption{Slice-level counts}
        \label{subtab:slice_counts_resizebox}
        \begin{tabular}{lcrrr}
          \toprule
          \textbf{Type} & \textbf{Source} & \textbf{COVID} & \textbf{non-COVID} & \textbf{Total} \\
          \midrule
          \multirow{4}{*}{Training}
            & 0 & 48,740 & 45,989 & 94,729 \\
            & 1 &  8,143 & 10,988 & 19,131 \\
            & 2 & 10,771 & 47,846 & 58,617 \\
            & 3 & 30,834 & 26,974 & 57,808 \\
          \midrule
          \multirow{4}{*}{Validation}
            & 0 & 10,957 & 11,909 & 22,866 \\
            & 1 &  2,070 &  2,814 &  4,884 \\
            & 2 &      0 & 13,283 & 13,283 \\
            & 3 &  7,739 &  8,942 & 16,681 \\
          \midrule
          \multirow{4}{*}{Total}
            & 0 & 59,697 & 57,898 & 117,595 \\
            & 1 & 10,213 & 13,802 &  24,015 \\
            & 2 & 10,771 & 61,129 &  71,900 \\
            & 3 & 38,573 & 35,916 &  74,489 \\
          \midrule
          Testing & -- & -- & -- & 306,203 \\
          \bottomrule
        \end{tabular}
      \end{subtable}
      
    \end{minipage}%
  }
\end{table}

\subsection{Experimental Setup and Details}

All experiments were conducted using the PyTorch framework on NVIDIA RTX 3090 GPUs. Implementation details will be introduced as follows:

\textbf{Model Architecture:} We compare two architectures in our experiments: EfficientNet-B7 \cite{efficientnet,hsu2024} and Swin Transformer-Base \cite{swin}. The EfficientNet model is based on the EfficientNet-B7 backbone with a classification head, while the Swin Transformer uses the base configuration with hierarchical feature learning. Both models process the 8 selected representative slices from each CT scan to make final predictions through global average pooling and fully connected classification layers.

\textbf{Training Configuration:} We employ EfficientNet-B7 as our backbone architecture. The model is trained using the Adam optimizer with a learning rate of $1 \times 10^{-4}$ and weight decay of $5 \times 10^{-4}$. The training batch size is set to 20, validation batch size to 1, and training is performed for up to 3 epochs with early stopping based on validation performance. We use binary cross-entropy as the loss function with mixed precision training (AMP) for computational efficiency. All input images are resized to $256 \times 256$ pixels and normalized using ImageNet statistics.

\textbf{Data Augmentation:} To improve generalization and robustness, we apply several data augmentation techniques during training:
\begin{itemize}
    \item Horizontal flipping with 50\% probability
    \item Shift-scale-rotate transformation with shift limit 0.2, scale limit 0.2, and rotation within ±30 degrees (50\% probability)
    \item Hue, saturation, and value adjustment with shift limits of ±0.2 (50\% probability)
    \item Random brightness and contrast adjustment within ±0.2 range (50\% probability)
    \item Coarse dropout with 20\% probability for regularization
    \item Normalization using ImageNet statistics \cite{IMAGENET}
\end{itemize}

For validation, we apply only resizing to $256 \times 256$ pixels and normalization to ensure consistent evaluation conditions.

\textbf{Cross-Validation Strategy:} We apply stratified 5-fold cross-validation for the training set. This approach maintains the class distribution of COVID-19 and non-COVID-19 cases within each fold, preserving the proportion of positive and negative samples as in the original dataset. In each cross-validation round, the training folds were combined with the validation set, resulting in a total of five folds per round.

\textbf{Evaluation Metrics:} We use F1-score as our primary evaluation metric, which provides a balanced measure of precision and recall. Additionally, we report AUC-ROC scores to assess the model's performance across different operating points.

%For the Swin Transformer, we use pre-trained weights from ImageNet and fine-tune all layers. The model uses a window size of 7×7 for self-attention computation and employs shifted window partitioning for computation across different scales.

%Our preprocessing pipeline reduces data redundancy by approximately 70\% while processing the diagnostic information across different medical centers.

\subsection{Experiment Results}
We evaluate the performance of both EfficientNet-B7 and Swin Transformer models under two different training strategies: source-specific training and multi-source joint training.

\textbf{Source-Specific Training:} Table~\ref{tab:source_specific_performance} presents the performance when each model is trained separately on individual sources and evaluated on the corresponding validation sets. The EfficientNet-B7 model achieves F1-scores of 97.53 and 90.90 on Sources 0 and 3, respectively. The Swin Transformer achieves F1-scores of 93.17 and 94.15 on Sources 0 and 3. Both models show lower performance on Source 2, where the validation set contains only positive samples.

\textbf{Multi-Source Joint Training:} When training on all sources combined, as shown in Table~\ref{tab:overall_performance}, the EfficientNet-B7 model achieves an overall F1-score of 94.68, compared to the Swin Transformer's 93.34. The EfficientNet-B7 model achieves an AUC-ROC of 0.9813 compared to 0.9797 for the Swin Transformer.

The lower performance on Source 2 for both models can be attributed to the validation set containing only positive samples (COVID cases), which affects the calculation of binary classification metrics. This indicates the importance of balanced validation sets for performance evaluation in multi-source medical imaging tasks.
\begin{table}[!ht]
\centering
\caption{Performance comparison across different hospitals between the proposed EfficentNet-B7 and Swin Transformer on the COVID-19-CT-DB validation set.}
\label{tab:source_specific_performance}
\scalebox{0.9}{%
\begin{tabular}{lccc}
\toprule
Model type & Hospital & AUC ROC & F1-score \\
\midrule
\multirow{4}{*}{Swin Transformer}
& 0 & 0.9922 & 93.17 \\
& 1 & 0.9515 & 87.17 \\
& 2 & NA$^{*}$ & 100$^{*}$ \\
& 3 & 0.9698 & 94.15 \\
\midrule
\multirow{4}{*}{EfficentNet-B7}
& 0 & 0.9981 & 97.53 \\
& 1 & 0.9564 & 87.96 \\
& 2 & NA$^{*}$ & 49.21$^{*}$ \\
& 3 & 0.9672 & 90.90 \\
\bottomrule
\end{tabular}%
}

\vspace{0.2cm}
\small
$^{*}$ Results affected by class imbalance (validation set contains only covid samples).
\end{table}
\begin{table}[!ht]
\centering
\caption{Overall performance comparison between the proposed EfficentNet-B7 and Swin Transformer on the COVID-19-CT-DB validation set (multi-source joint training).}
\label{tab:overall_performance}
\begin{tabular}{lcc}
\toprule
Model type & AUC ROC & F1-score \\
\midrule
Swin Transformer & 0.9797 & 93.34 \\
EfficentNet-B7 & 0.9813 & 94.68 \\
\bottomrule
\end{tabular}
\end{table}

Our results show several findings: (1) The EfficientNet-B7 model outperforms the Swin Transformer in our evaluation scenarios, (2) Both models show performance variation across sources, with lower scores on Source 2, and (3) The class imbalance in Source 2's validation set affects performance evaluation, highlighting considerations for dataset balance in multi-institutional studies.
\section{Conclusion}

We present a robust solution for the Multi-Source COVID-19 Detection Challenge using the SSFL framework with KDS strategy. Our EfficientNet model achieves an F1-score of 94.68, outperforming the Swin Transformer's 93.34, while reducing data redundancy by 70\%. The results demonstrate that well-designed 2D architectures with effective preprocessing can achieve superior performance compared to complex transformer-based approaches, providing a practical foundation for AI-assisted COVID-19 detection in clinical settings.

{
    \small
    \bibliographystyle{ieeenat_fullname}
    \bibliography{main}
}


\end{document}